\documentclass[a4paper,fleqn,,useAMS]{mnras}
\usepackage{newtxtext,newtxmath}
\usepackage[T1]{fontenc}
\usepackage{ae,aecompl}
\usepackage{graphicx}	
\usepackage{amsmath}	
\usepackage{amssymb}	
\usepackage{multicol}        

\title[Surface habitability of Kepler-452b]{Quantitative estimates of the surface habitability of Kepler-452b}

\author[L. Silva et al.]{
Laura Silva,$^{1}$\thanks{Contact e-mail: \href{mailto:silva@oats.inaf.it}{silva@oats.inaf.it}}
Giovanni Vladilo,$^{1}$
Giuseppe Murante$^{1}$
and Antonello Provenzale$^{2}$
\\
$^{1}$National Institute for Astrophysics, INAF-OATs, Trieste, Italy\\
$^{2}$Institute of Geosciences and Earth Resources, CNR, Pisa, Italy
}

\date{Accepted XXX. Received YYY; in original form ZZZ}
\pubyear{2017}

\begin{document}
\label{firstpage}
\pagerange{\pageref{firstpage}--\pageref{lastpage}}
\maketitle

\begin{abstract}
Kepler-452b is currently the best example of an Earth-size planet in the habitable zone of a sun-like star, a type of
planet whose number of detections is expected to increase in the future. Searching for biosignatures in the supposedly thin atmospheres of these planets is a challenging goal that requires a careful selection of the targets.   
Under the assumption of a rocky-dominated nature for Kepler-452b, we considered it as a test case to calculate a temperature-dependent habitability index, $h_{050}$,  designed to maximize the potential presence of biosignature-producing activity (Silva et al.\ 2016). 
The surface temperature has been computed for a broad range of  climate factors using a climate model designed for terrestrial-type exoplanets (Vladilo et al.\ 2015). After fixing the planetary data according to the experimental results (Jenkins et al.\ 2015), we changed the surface gravity, CO$_2$ abundance, surface pressure, orbital eccentricity, rotation period, axis obliquity and ocean fraction within the range of validity of our model. For most choices of parameters we find habitable solutions with $h_{050}>0.2$ only for  CO$_2$ partial pressure $p_\mathrm{CO_2} \lesssim 0.04$\,bar. At this limiting value of CO$_2$ abundance the planet is still habitable if the total pressure is $p \lesssim 2$\,bar. In all cases the habitability  drops for eccentricity $e \gtrsim 0.3$.
Changes of rotation period and obliquity affect the habitability  through their impact on the equator-pole temperature difference rather than on the mean global temperature. We calculated the variation of $h_{050}$ resulting from the luminosity evolution of the host star for a wide range of input parameters. Only a small combination of parameters yield habitability-weighted lifetimes $\gtrsim 2$\,Gyr,  sufficiently long to develop atmospheric biosignatures still detectable at the present time.
\end{abstract}

\begin{keywords}
astrobiology -- planets and satellites: individual: Kepler-452b -- planets and satellites: terrestrial planets 
\end{keywords}

\section{Introduction}

One of the drivers of exoplanet research is the quest for planets that host a remotely detectable biosphere. This endeavor requires a preliminary identification of planets that have the capability of hosting life on their surface over astronomical/geological time scales. At variance with life confined under the surface,  surface life can generate atmospheric biosignatures detectable by means of exoplanet atmospheric spectroscopy (e.g. Tinetti, Encrenaz \& Coustenis 2013; Kasting et al. 2014; Seager 2014). The persistence of habitability conditions over gigayear time scales 
increases the probability of detecting a diffuse biosphere. The identification of planets with conditions of long-term surface habitability is not trivial, partly because the habitability is influenced by many factors (e.g. Seager 2013; G{\"u}del et al. 2014). However, a preliminary identification can be obtained by searching for terrestrial type exoplanets lying in the habitable zone (Kasting 1988; Kasting, Whitmire \& Reynolds 1993; Kopparapu et al. 2013, 2014) of late-type stars for a large fraction of the stellar lifetime. Once an exoplanet with such characteristics is found, a detailed study of its properties is required to understand which are the exact conditions that would maximize the detectability of a potential biosphere. 

So far, only a small fraction of known exoplanets lies in the habitable zone (HZ) and, at the same time, is of terrestrial type\footnote{ see e.g. http://phl.upr.edu/projects/habitable-exoplanets-catalog}. Most of these planets have been found around M-type stars, partly because the semi-amplitude of radial velocity curves and the depth of transit light curves become stronger with decreasing mass and radius of the host star. However, planets in the habitable zone of low-mass stars are subject to a number of life-threatening challenges, such as the impact of nearby, intense stellar eruptions and the potentially harsh climate conditions resulting from the synchronization of the spin-orbital periods (e.g. Shields, Ballard \& Johnson 2016). Whether such conditions may allow the presence of a widespread, remotely detectable biosphere remains to be demonstrated. Terrestrial type planets in the HZ of solar-type stars do not suffer from these challenges and therefore are more promising as far as the effective capacity of generating biosignatures is concerned. These planets are particularly difficult to detect and only a few cases have been discovered so far. 

At the present time, the best example is Kepler-452b, a terrestrial-sized planet  in the HZ of a solar-type central star (Jenkins et al. 2015). The planetary mass is not currently measured. The probability of a pure rocky composition for this planet has been quantified by Jenkins et al.\  to lie between $49$\% and $62$\%. A rock-ice mixture could extend the probability that the planet is rocky- rather than gaseous-dominated. On the other hand, its measured radius, $60$\% larger than the Earth radius, lies in the super-Earths to sub-Neptunes transition region, where a substantial volatile envelope could be contributing to the observed size. 
(e.g., Lopez \& Fortney 2014;  Marcy et al. 2014; Rogers 2015).
Therefore, at this stage any inference on the habitability of Kepler-452b requires to assume its rocky nature.  

By comparing the evolution of the insolation flux of Kepler-452b with the insolation limits of the HZ of solar-type stars, Jenkins et al. (2015) conclude that Kepler-452b has likely been in the HZ for the past $\sim 6$\,Gyr and should remain there for another $\sim 3$ Gyr. 
Given the unique conditions of Kepler-452b, a detailed modelization of its climate is required to assess its habitability taking into account a full set of climate factors and not just the (evolution of its) insolation.   

As in any study of planetary climate, the use of a hierarchy of models is fundamental to cast light on different aspects of the climate of Kepler-452b. Recently, a coupled atmosphere-ocean GCM with three atmospheric compositions has been used to simulate the climate of Kepler-452b (Hu et al.\ 2017). The results of that work indicate that Kepler-452b is habitable if the CO$_2$ concentration in its atmosphere is comparable or lower than that in the present-day Earth atmosphere. In addition to the atmospheric composition, however, there are several  factors unconstrained by observations which may significantly affect the climate of Kepler-452b and have not been considered so far. These factors include orbital eccentricity, planet surface gravity, total surface pressure, rotation period, axis inclination, and ocean coverage. GCM simulations are not suitable to perform a systematic study of all these factors because they require a large amount of computing resources. In the present work we investigate the impact of these factors on the habitability of Kepler-452b using the ESTM (Earth-like planet Surface Temperature Model), a flexible model developed for the study of the surface temperature of Earth-like exoplanets (Vladilo et al.\ 2015, V15 hereafter). In addition, at variance with previous work, we provide a quantitative estimate of the fraction of habitable surface of Kepler-452b for each set of planetary parameters considered. The quantitative estimate is calculated with a temperature-dependent index of habitability  designed to maximize the potential presence of complex life and atmospheric biosignatures (Silva et al. 2016). The evolution of stellar luminosity, in conjunction with variations of other climate factors, is also studied to understand the time span of the habitability of this planet. In the next section we briefly explain our methodology and test our model using the results presented by Hu et al. (2017). In Section \ref{secResults} we describe the impact of the climate factors on the present-day habitability of 
Kepler-452b. In Section \ref{secEvoHab} we apply our methodology to explore the  evolution of the habitability of Kepler-452b resulting from the gradual rise of the luminosity of its host star. The results are discussed and summarized in Section \ref{secConc}.

\section{Method}\label{secMethod}

The ESTM is an upgraded type of seasonal and latitudinal energy balance model (EBM) featuring a multi-parameter description of the  planet  surface and of the meridional and vertical  energy transport.
We provide here a summary of its main features and equations, and refer to V15 for more details.

\begin{description}

\item[-] Diffusion equation for zonal energy balance:

We adopt the commonly used diffusion equation of energy balance (e.g., North \& Coackley 1979; North et al.\ 1983; Williams \&  Kasting 1997; Spiegel, Menou \& Scharf 2008, 2009), with which
the vertical energy balance is computed at each latitude, and the horizontal heat transport is treated as a diffusion process:

\begin{equation}
C\,\frac{\upartial T}{\upartial t} - \frac{\upartial}{\upartial x} \left [D \left ( 1-x^2  \right )  \frac{\upartial T}{\upartial x}  \right ] + I = S \, (1-A)
\label{eq:ebm}
\end{equation}

where $x=\sin \varphi$, with $\varphi$ the latitude. The equation is solved for $T=T(\varphi,t_s)$, the latitude and seasonal ($t_s$)-dependent surface temperature. 
All terms depend, directly or through the temperature, on time and latitude, and are per unit surface.

The terms for the vertical energy transport are $I$, the outgoing longwavelength thermal radiation (OLR); $S$, the incoming stellar radiation; $A$, the top of atmosphere planetary albedo. The horizontal heat transport is given by the second term of on the left hand side of the equation, the diffusion term, that represents the amount of heat per unit time and area that is exchanged along the meridional direction.   
$C$ is the zonal heat capacity that drives the thermal inertia of the system. 

The latitude and seasonal dependence of $S$ is exactly computed as a function of the planetary inclination and orbital parameters. The accounting of longitude-averaged quantities implies that the time dependence refers to the orbital motion, and that the model can be applied only to planets with a rotation period much smaller than the orbital period. The possibility to model planets with a range of characteristics rely on a physically based parametrization of the different terms of Eq.\ \ref{eq:ebm}, as summarized below (see Section 2 and Figure 1 in V15).\\

\item[-] Horizontal transport:

The meridional transport introduced in V15 is modelled with physically-based algorithms validated with 3D climate models, that mimic the extratropical transport of a rapidly rotating planet (Section 2.1 in V15). The starting point is the definition of the coefficient $D$ of the diffusion term in Eq. \ref{eq:ebm}. In analogy with heat diffusion, it is defined by the following relation: 

\begin{equation}  
\Phi \equiv -D \frac{\upartial T}{\upartial \varphi}
\label{eq:phi}
\end{equation}  
 
where $2 \pi R^2 \Phi \cos\varphi$ is the net rate of atmospheric energy transport across a circle of given latitude. 
$\Phi$ can be expressed as the zonal flux of moist static energy of the atmosphere. By adopting the analytical treatment of the baroclinic circulation proposed and tested with GCM results by Barry et al. (2002), V15 derived $\Phi$ and therefore $D$ parametrized as a function of planetary properties implied in the horizontal energy transport: 
planet radius, surface atmospheric pressure, surface gravitational acceleration, planet angular velocity, and relative humidity. The smoothing of the meridional temperature gradient typical of the tropical regions inside the Hadley cell is simulated with a dedicated algorithm aimed to enhance $D$ in correspondence of the thermal equator. The functional dependences of this modelling are tested agaist the GCM model by Kaspi \& Showman (2014; see Figure 3 in V15).  \\

\item[-] Vertical transport: 

The vertical transport is solved by making use of radiative-convective atmospheric column calculations. 

In the current version of the ESTM, these calculations are performed with standard radiation codes developed at the National Center for Atmospheric Research (NCAR), as part of the Community Climate Model (CCM) project (Kiehl et al.\ 1998). In practice, the  OLR and top-of-atmosphere albedo $A$ are calculated in clear sky conditions and tabulated as a function of surface temperature, pressure and surface albedo for a given set of surface gravity, relative humidity, and atmospheric composition. 
With these radiative models, the atmospheric content of CO$_2$ and CH$_4$ can be varied as long as these gases remain in trace amounts in an otherwise Earth-like atmospheric composition. \\

\item[-] Parametrization of planetary surface: 

The planet surface is parameterized with the coverage, thermal capacity and albedo of oceans, continents and ice, which are tuned with Earth experimental data. 
A schematic geography can be specified by assigning a constant or zonal-dependent coverage of oceans $f_o$, a free parameter. This determines the fraction of continents.
The zonal coverage of ice is calculated with an algorithm that depends on the temperature and the fraction of time during which the  temperature is below the water freezing point. 
The zonal coverage of water, continent and ice determines the surface thermal capacity $C$ by averaging over the thermal capacity 
of each type of surface and its coverage. The ocean albedo makes use of a physical recipe which is a function of stellar zenith distance.
The radiative effects of the clouds are accounted for at a later stage by means of a parameterization for their surface coverage, albedo and infrared absorption (details in Section 2.3 of V15).

\end{description}

Given a set of planetary parameters and a set of initial values, the ESTM simulations iteratively search for a stationary solution of the surface temperature $T(\varphi,t_s)$.
Full details on the ESTM and its limits of validity are given in Vladilo et al. (2015). 

The solution $T(\varphi,t_s)$  is used to calculate several parameters relevant to the habitability of the planet surface, such as the mean global orbital temperature, $<\!T\!>$, the mean orbital temperature difference between the equator and the poles, $\Delta T_\mathrm{ep}$, and the mean orbital coverage of ice, $f_\mathrm{ice}$. In addition, the fractional habitability of the planet surface is calculated from the solution $T(\varphi,t_s)$ by adopting two temperature thresholds, $T_1$ and $T_2$, representative of the thermal limits of a habitable environment. In practice, we define a boxcar function $H(\varphi,t_s)$ such that 
$H(\varphi,t_s)\equiv 1$ when $T(\varphi,t_s) \in [T_1,T_2]$ and $H(\varphi,t_s)\equiv 0$ when $T(\varphi,t_s) \notin [T_1,T_2]$. We then integrate $H(\varphi,t_s)$ in latitude  and time. The integration in $\varphi$ is weighted according to the area of each zone. The integration in $t_s$ is performed over one orbital period. From this double integration we obtain an index of habitability, $h$, which represents the global and orbital mean fraction of planet surface that satisfies the assigned temperature limits.  Besides the mean global temperature  $<\!T\!>$, also $\Delta T_\mathrm{ep}$ and $f_\mathrm{ice}$ help to understand why a specific set of climate factors may yield a low or high value of habitability $h$. For instance, if the equator-pole gradient $\Delta T_\mathrm{ep}$  significantly exceeds the temperature difference $(T_2-T_1)$, the habitability  is necessarily low.  Also the ice coverage $f_\mathrm{ice}$ plays a key role in determining $h$ if $T_1$ is set equal to the melting point of water ice. 

The habitability index can be defined in different ways, depending on the choice of the temperature limits $T_1$ and $T_2$. The water melting point, $T_\mathrm{ice}$, and boiling point,  $T_\mathrm{vapor}$, provide pressure-dependent limits  that can be used to calculate a liquid water index of habitability, $h_\mathrm{lw}$ (Vladilo et al. 2013, 2015). Alternatively, the biological limits $[T_1,T_2] \equiv [0^{\circ}\mathrm{C},50^{\circ}\mathrm{C}]$, typical of multicellular poikilotherms with active metabolism and capability of reproduction (Precht et al. 1973, Clarke 2014), can be used to calculate an index $h_{050}$ based on the thermal response of life (Silva et al. 2016). The index $h_{050}$ offers two advantages with respect to the liquid water index $h_\mathrm{lw}$. First, the photosynthetic production of atmospheric biosignatures is maximized in the interval $[0^{\circ}\mathrm{C},50^{\circ}\mathrm{C}]$, meaning that exoplanets with high values of $h_{050}$ are optimal targets to search for biosignatures in planetary atmospheres. Second, the index $h_{050}$ is weakly affected by the uncertainties inherent to the modelization of hot, moist atmospheres, which instead require specific climate models (e.g. Leconte et al. 2013). In particular, the upper limit $T_2=50^{\circ}\mathrm{C}$ is sufficiently low to prevent the onset  of the runaway greenhouse instability, which cannot be tracked with our simplified model. For the above reasons, in the rest of this work we adopt the index $h_{050}$. More details on the thermal limits of life and examples of habitable zones calculated with the $h_\mathrm{lw}$ and $h_{050}$ indices can be found in Silva et al. (2016). 


\begin{table*}
  \centering
  \caption{Comparison with the results of Hu et al. (2017)}
  \label{tabValidation}
  \begin{tabular}{lcccccl}
    \hline
    Case &  Model  &  $<T>(K)$  &  $T_\mathrm{trop}(K)$  &  $T_\mathrm{poles}(K)$  &  Ice limit  &  Reference \\
    \hline
    E-CO$_2$ & GCM & 293 & 310 & 240 & 55$^\circ$ & Hu et al. (2017) \\
                 & ESTM & 297 & 317 & 245 & 58$^\circ$ & This work \\
    \hline
    L-CO$_2$ & GCM & 285 & 308 & 230 & 45$^\circ$ & Hu et al. (2017) \\
                & ESTM & 278  & 305  &  219 & 42$^\circ$ & This work \\
    \hline
  \end{tabular}
\end{table*}

\section{Results}\label{secResults}

To simulate the climate of Kepler-452b we adopted the measurements of stellar, orbital, and planetary data obtained from the {\em Kepler} mission observations (Jenkins et al. 2015). In particular, we adopted the stellar luminosity $L_\star = 1.21 L_\odot$, stellar mass $M_\star = 1.037 M_\odot$, orbital period, $P_\mathrm{orb}= 384.8$\,d (Earth days), and planet radius $R=1.63\,R_\oplus$. 
The resulting insolation is 10\% higher than the present-day insolation of the Earth, for a zero eccentricity orbit.
Based on these data, we performed a comparison with the Earth climate and a validation test with a 3D climate model of Kepler-452b. We then run several series of ESTM simulations aimed at exploring the impact of multiple climate factors on the habitability of Kepler-452b.

\subsection{Comparison with the Earth climate}\label{secCompEarth}

Among the climate factors measured with the observations of Kepler-452b, the insolation $S=1.10\,S_\oplus$ and radius $R=1.63\,R_\oplus$ represent two major differences relative to the case of the Earth. Quite interestingly, the larger insolation and the larger radius have opposite effects on the mean equator-pole gradient $\Delta T_\mathrm{ep}$, a critical parameter for habitability. The larger insolation rises the efficiency of the moist transport and decreases $\Delta T_\mathrm{ep}$. On the other hand, the larger distance between the equator and the poles decreases the fraction of energy delivered to the polar regions and increases $\Delta T_\mathrm{ep}$. Both effects have been tested with 3D models (e.g., Kaspi \& Showman 2015) and are incorporated in the ESTM (Vladilo et al. 2015). To quantify the combined impact of both factors, we increased the solar luminosity by 10\% and adopted a planetary radius $R=1.63\,R_\oplus$ 
in the reference Earth model of the ESTM, leaving unchanged all the remaining parameters. The higher insolation yields a mean global annual temperature $<\!T\!>=307.4$\,K, significantly higher than that of the Earth,  $<\!T\!>=289.4$\,K, but still well within the interval $[T_1,T_2]=$[273\,K,323\,K].  In fac the resulting habitability, $h_{050}=0.99$, is higher than that of the Earth, $h_{050}=0.87$. This is due to two factors. First, the concurrent change of insolation and radius yields a gradient $\Delta T_\mathrm{ep} =43.6$\,K $< (T_2-T_1)$, very similar to that of the Earth, $\Delta T_\mathrm{ep}=41.6$\,K. Second, due to the higher temperatures, the polar coverage of ices disappears, leading to a higher fraction of habitable surface. 
This result is interesting for illustrative purposes, but is unlikely to reflect the actual conditions of Kepler-452b because, in addition to radius and insolation, also other climate factors of this planet are likely different from the case of Earth, as we discuss below.

\subsection{Validation test}

To perform a validation test with an independent climate model of Kepler-452b, we used the results obtained from the coupled atmosphere-ocean GCM of Hu et al. (2017). For the sake of consistency, we adopted 
an orbital eccentricity $e=0$, planet rotation period $P_\mathrm{rot}=1$\,d, axis tilt $\epsilon=0^\circ$, surface gravitational acceleration $g=1.6\,g_\oplus$, surface atmospheric pressure $p=1$\,bar, and 100\% ocean coverage. To compare the cases E-CO$_2$ and L-CO$_2$ considered by Hu et al. (2017), we adopted CO$_2$ concentrations of 355 ppmv and 5 ppmv, respectively. The resulting values of  mean global orbital temperature, $<\!T\!>$, tropical temperature, $T_\mathrm{trop}$, polar temperature, $T_\mathrm{poles}$, and ice limit are compared in Table \ref{tabValidation}. In the case E-CO$_2$ the temperature predictions of the GCM and ESTM are reproduced within $\sim 5$\,K and the ice limit within a few degrees in latitude. In the case L-CO$_2$ the temperature differences are slightly larger but  the ice limits are still very close. Considering the wide differences in the climate models adopted, the agreement is encouraging. The general agreement of the extremes of zonal temperature and of the latitudinal extension of the ice cover suggests that the temperature-dependent habitability index that we adopt would yield very similar results with both models.

\begin{table*}
\centering
\caption{Reference models of Kepler-452b$^a$}
\label{tabModels}
\begin{tabular}{cccccccccl}
\hline
Model  &   $M/M_\oplus$   &   $g/g_\oplus$   &   $p$CO$_2$\,(ppmv)   &   $p^b$\,(bar)   &
  $e$  &  $P_\mathrm{rot}$\,(d) & $\epsilon\,(^\circ)$   &   $f_o$   &  Comment \\
\hline
RL &  4.3 & 1.6 & 10 & 2.6   & 0.0  & 1.0 & 0 & 1.0 & Rocky, low CO$_2$  \\
RE &  4.3 & 1.6 & 380 & 2.6   & 0.0  & 1.0 & 0 & 1.0 & Rocky, Earth-like CO$_2$\\
RH &  4.3 & 1.6 & 38000 & 2.6  & 0.0  & 1.0 & 0 & 1.0 & Rocky, high CO$_2$\\
OL &  2.7 & 1.0 & 10 & 1.0   & 0.0   & 1.0 & 0 & 1.0  & Rocky/water, low CO$_2$\\
OE &  2.7 & 1.0 & 380 & 1.0   & 0.0 & 1.0 & 0 & 1.0  & Rocky/water, Earth-like CO$_2$\\
OH &  2.7 & 1.0 & 38000 & 1.0   & 0.0  & 1.0 & 0 & 1.0 & Rocky/water, high CO$_2$\\
\hline
\multicolumn{10}{l}{$^a$ For each model the surface pressure, orbital eccentricity, rotation period, axis tilt and ocean fraction
have been varied in the intervals: }\\
\multicolumn{10}{l}{ $0.3 \leq p \mathrm{(bar)} \leq 5$, $0 \leq e \leq 0.5$, $0.5 \leq P_\mathrm{rot} \mathrm{(d)} \leq 2.0$, $0^\circ \leq \epsilon \leq 45^\circ$, and $0.1 \leq f_o \leq 1$, respectively. }\\
\multicolumn{10}{l}{ These parameters are varied only one at a time in each series of simulations, fixing the others
to the reference values listed in the table.}\\
\multicolumn{10}{l}{$^b$ Educated guess of surface atmospheric pressure obtained from Eq. (\ref{scalingLaw}).}\\
\end{tabular}
\end{table*}

\subsection{Planetary models of Kepler-452b}

To perform a large series of ESTM climate experiments of Kepler-452b we introduced a set of planetary models  aimed at exploring the parameter space that is unconstrained by the observations.  In each model we adopted a fixed value of surface gravity acceleration, $g$, and CO$_2$ content in the atmosphere, $p$CO$_2$. As we explain below, we considered two possible values of $g$ and three values of $p$CO$_2$, giving a total of six models shown in Table \ref{tabModels}. We also selected with care the value of surface atmospheric pressure, $p$. The model parameters $g$ and $p$ are particularly important because they determine the columnar mass of the planetary atmosphere, $N_\mathrm{col}=p/g$, which is one of the most critical climate factors not measurable at present time in terrestrial-type exoplanets. Indeed, the atmospheric columnar mass governs the efficiency of the horizontal energy transport and, in conjunction with the atmospheric composition, also the efficiency of the vertical radiative transport. To model  $N_\mathrm{col}$ we adopted a plausible set of $p$ and $g$ values. 

To obtain an estimate of $g$, we combined the measurement of the planet radius, $R=1.63\,R_\oplus$, with an assumption on the internal composition of Kepler-452b consistent with the observational constraints. According to Jenkins et al. (2015), the probability that Kepler-452b
is rocky, with small or negligible iron core, lies between 49\% and 62\%. For consistency with their analysis, we adopted a rock mass fraction of 1 in the mass-radius relation (8) of Fortney, Marley \& Barnes (2007), corresponding to a planet of pure silicates. For $R=1.63\,R_\oplus$,
that mass-radius relation yields a mass $M=4.3\,M_\oplus$ and a surface gravity $g=1.6\,g_\oplus$ that we adopt as a reference for the modes RL, RE, and RH in Table \ref{tabModels}. According to Jenkins et al. (2015), the probability that Kepler-452b has an Earth-like rock-iron mixing ratio of 2/3, or a denser composition, is significantly smaller (16\% to 22\%). This unlikely internal composition was not considered in our models. To assess the impact of the uncertainty of the $g$ parameter, we instead considered the possibility that Kepler-452b  has a mean density lower than that of pure silicates. Specifically, we considered the case $R=1.63\,R_\oplus$ and $g=1.0\,g_\oplus$ which, according to the mass-radius relation (7) in Fortney et al. (2007), corresponds to a rocky planet with an ice mass fraction of 0.25 and total mass $M=2.7\,M_\oplus$. We adopted these value of $g$ and $M$ for the models OL, OE, and OH. 

For each value of $g$ and $M$ we estimated a reference value of $p$ as follows.
We started from the hydrostatic equilibrium condition,  $p=g N_\mathrm{col}$, and the relation
$N_\mathrm{col} \cong M_\mathrm{atm}/(4\pi R^2)$, where $M_\mathrm{atm}$ is the mass of the planet atmosphere. 

Combining these expressions with the scaling relation $g \propto M/R^2$, one has $p \propto  M_\mathrm{atm} \, M / R^4$. To derive $p$ from this expression we need an estimate of $M_\mathrm{atm}$. The atmospheric mass is hard to predict because it depends on the history of the planet atmosphere, which starts from the acquisition of volatiles at the stage of planet formation and may continue with  episodes of late delivery of volatiles, atmospheric loss, and mass exchanges with the surface. For planets more massive than the Earth, as it is the case of Kepler-452b, we may assume that they are capable of retaining their atmosphere.


The best we can do for the other mechanisms is to assume that (1) starting from the last stages of planet formation, the amount of volatiles acquired by the planet is proportional to planet mass, (2) the fraction of atmospheric mass exchanged with the surface is comparable in different planets. This second assumption is reasonable for planets with surfaces undergoing similar geophysical and geochemical processes. With these assumptions, $M_\mathrm{atm} \propto M$ and we obtain the 
scaling relation\footnote{The same relation was adopted by Kopparapu et al. (2014).}
\begin{equation}
p \propto { M^2 \over R^4} ~
\label{scalingLaw}
\end{equation}
that we used to estimate the representative values of $p$ shown in Table \ref{tabModels}. Given the uncertainty of this derivation, we also explored the effect of variations of $p$ in a broader interval (see note $a$ in Table \ref{tabModels}). 

To test the impact of variations of atmospheric composition, we experimented three different levels of non-condensable greenhouse gas by changing the content of CO$_2$ in an otherwise Earth-like composition. Specifically, a representative value for the present-day Earth, $p$(CO$_2$)=380 ppmv, was adopted in models RE and OE. The minimum CO$_2$ content for plants with C4 photosynthesis systems, $p$(CO$_2$)=10 ppmv (Caldeira \& Kasting 1992), was adopted
for the models RL and OL. Finally, a CO$_2$ content 100 times higher than that of the present Earth, i.e.  $p$(CO$_2$)=38000 ppmv, was adopted for the models RH and OH.  

For the remaining parameters, we explored a range of values of orbital eccentricity, $e$, planet rotation period, $P_\mathrm{rot}$, obliquity of the rotation axis, $\epsilon$, and fraction of surface covered by oceans, $f_o$ (see note $a$ of Table \ref{tabModels}). For each one of the six models we performed a series of climate experiments varying $p$ or one of these parameters while fixing the others to the reference values shown in Table \ref{tabModels}.

\begin{figure*}     
\includegraphics[width=\linewidth]{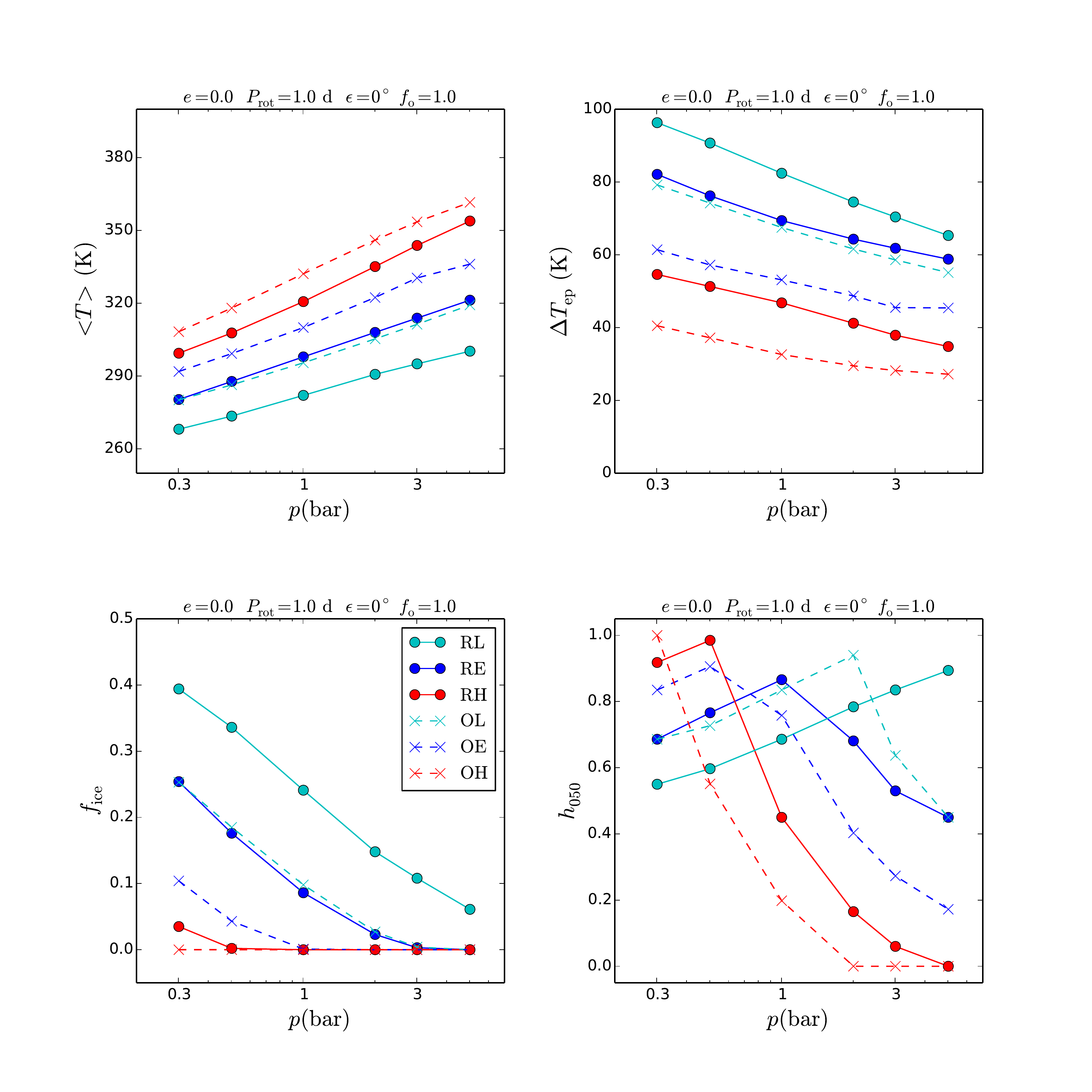} 
\caption{
Impact of surface atmospheric pressure, $p$, on the surface climate and habitability of Kepler-452b for the six planetary models listed in Table \ref{tabModels}. The other parameters were fixed as indicated in the title of each panel (see table note in Table \ref{tabModels}).
 Upper  panels: mean global orbital temperature, $<\!T\!>$, and mean orbital equator-pole temperature difference, $\Delta T_\mathrm{ep}$. Lower panels: ice coverage, $f_\mathrm{ice}$, and habitability index $h_{050}$. 
}
\label{figPressure}
\end{figure*}

\begin{figure*}     
\begin{center}
\includegraphics[width=\linewidth]{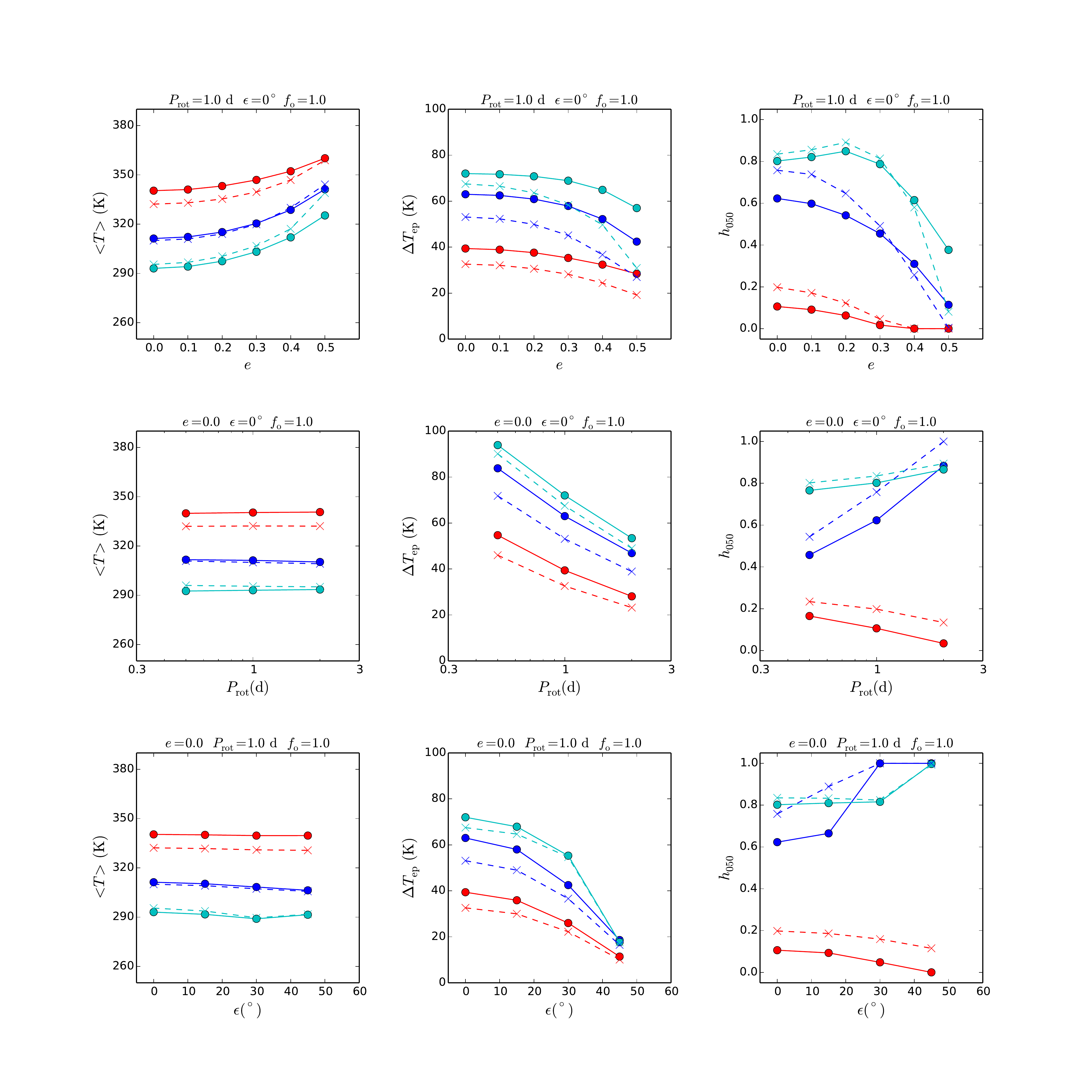} 
\caption{
Impact of orbital eccentricity, $e$, planet rotation period, $P_\mathrm{rot}$, and axis obliquity, $\epsilon$, on the surface temperature and habitability of Kepler-452b for the six planetary models listed in Table \ref{tabModels} (see legend in Fig. \ref{figPressure}). The surface pressure of the rocky planet models (solid curves) and ocean/rocky models (dashed curves) was fixed to the reference values $p=2.6$\,bar and $p=1.0$\,bar, respectively (see Table \ref{tabModels}). The other parameters were kept fixed as indicated in the title of each panel (see also note in Table\ref{tabModels}). 
}
\label{figEccRotObl}%
\end{center}
\end{figure*}

\subsection{The present-day habitability of Kepler-452b}\label{secPreHab}

For studying the habitability of Kepler-452b at the present time, we adopted the present-day luminosity of the central star, which yields an insolation 10\% larger than that currently received by the Earth. 
Given the critical role of the atmospheric columnar mass $p/g$, and considering the uncertainty in the derivation of $p$ with Eq. (\ref{scalingLaw}), we first investigated the impact of variations of $p$, which have not been considered in previous studies of Kepler-452b.   

Fig. \ref{figPressure}  shows the impact of a change of surface atmospheric pressure on the habitability  of Kepler-452b. The six curves in each panel correspond to the models of Table \ref{tabModels}. Each curve was calculated by fixing the other parameters (see figure caption). In the top left panel one can see that the mean global orbital temperature, $<\!T\!>$, increases with $p$. This is due to the rise of the greenhouse effect with increasing atmospheric columnar mass, $N_\mathrm{atm}=p/g$, at constant atmospheric composition. In the second panel one can see that the mean orbital equator-to-pole temperature difference, $\Delta T_\mathrm{ep}$, decreases with increasing $p$. This is due to the rise of the meridional transport efficiency with increasing $N_\mathrm{atm}$. Due to the rise of global temperature and the decrease of meridional gradient, the mean global ice cover, $f_\mathrm{ice}$,  decreases with increasing pressure, as shown in the bottom left panel. The behaviour of the habitability index $h_{050}$, shown in the bottom right panel, can be easily understood in terms of the above effects: at low pressure, the habitability increases with $p$ because the ice coverage decreases; then, when the ice coverage disappears, the habitability decreases due to the rise of surface area with critical temperature $T>50^\circ$C. Habitable solutions are found in a broad range of surface pressure when the CO$_2$ content is Earth-like (blue curves) or smaller (green curves). Instead, the habitability drops very fastly with increasing $p$ when the CO$_2$ content is higher (red curves). The models OE, OL, and OH, with lower planet gravity (dashed curves), are somewhat warmer than the corresponding models RE, RL, and RH, representative of a pure rocky planet with higher gravity (solid curves). The warming is due to the higher atmospheric columnar mass, $N_\mathrm{atm}=p/g$, of models with lower $g$ at fixed $p$. As a result of the warming, the habitability may increase or decrease, depending on the presence of ice cover or not. The bottom right panel of Fig. \ref{figPressure} shows the importance of having a measurement or at least an information on $p$ in order to assess the planetary habitability.  In the case of the rocky planet models, the intersections of the habitability curves (solid lines) with the reference value of pressure ($p=2.6$\,bar; Table \ref{tabModels}), indicates that the planet is habitable only when the CO$_2$ content is equal or lower than the Earth's value. A similar result is found for the models with lower gravity by comparing the reference value ($p=1.0$\,bar) with the corresponding curves of habitability (dashed lines). In the rest of this discussion we adopt these reference values of $p$ to explore the impact of variations of other orbital and planetary parameters.  

In the top panels of Fig. \ref{figEccRotObl} we show the impact of variations of orbital eccentricity, $e$. A study of this effect is required because the orbital eccentricity of Kepler-452b is  poorly constrained by the observations (Jenkins et al. 2015). The mean insolation flux increases with eccentricity according to the relation $S \propto (1-e^2)^{-1/2}$ (see, e.g., Vladilo et al. 2013). In absence of a measurement of $e$, Jenkins et al. (2015) weighted this flux with a probability density for  eccentricity developed by Rowe et al. (2014) for multiple transiting planets. In this way, they derived a mean weighted insolation flux  1.17 $S_\oplus$ and concluded that uncertainties in the eccentricity do not significantly affect the habitability of Kepler-452b. However, the probability density for eccentricity of Rowe et al. (2014) may not be appropriate, because Kepler-452b is a single transiting planet and the distribution of eccentricities is broader for exoplanets that are not part of a multiple transit system (Xie et al. 2016). For this reason we re-assessed the potential impact of eccentricity on the habitability of Kepler-452b. With the ESTM we can estimate the fraction of habitable surface at each phase of the eccentric orbit, taking into account latitudinal variations of insolation and the impact of flux variations on the ice cover. The rise of global temperature with increasing $e$ calculated by the ESTM is visible in the top left panel of Fig. \ref{figEccRotObl}. This rise is accompanied by a moderate decrease of the equator-pole temperature gradient (central panel). The resulting index of habitability shows a sharp drop around $e \sim 0.3-0.4$ for all models (right panel). Before this drop, the models with minimum CO$_2$ content (green curves) show an initial rise of habitability with increasing $e$ due to the reduction of the ice cover. The (low) habitability of the models with high CO$_2$ content (red curves) vanishes completely at $e \sim 0.3$. This complex phenomenology is completely missed when the habitability is estimated only from the mean insolation flux.  

In the central row of panels in Fig. \ref{figEccRotObl} we show the impact of variations of rotation period, $P_\mathrm{rot}$, which so far have been not considered for Kepler-452b. Planetary rotation affects the horizontal transport on the planet surface due to the presence of Coriolis forces which tend to inhibit the transport from the tropics to the poles. At the moment, the observations are unable to constrain the rotation period of rocky exoplanets (for giants see e.g.\ Snellen et al.\ 2014).
Theoretical arguments may lead to constrasting predictions on the rotation properties of Kepler-452b.  On one side, the example of the Solar System shows that planets with larger masses rotate faster than the Earth, probably due to the conservation of angular momentum of a larger amount of material accreted at the stage of their formation. This argument would support a moderately high rotation speed since the mass of Kepler-452b is somewhat larger than that of the Earth. On the other side, the host star Kepler-452 is older than the Sun ($\sim 6$\,Gyr; Jenkins et al. 2015) and this gives sufficient time to neighbouring bodies, if any, to slow down the rotation period of Kepler-452b by means of tidal interactions. This tidal breaking scenario is currently not supported by observations because the planet is far from its central star and  no other planets have been found in its vicinity. However, the presence of undected neighbouring bodies, including exomoons with strong tidal effects, cannot be excluded. Given this uncertain scenario, we have explored the impact of variations of rotation period in the range $0.5 \lesssim P_\mathrm{rot}/\mathrm{d} \lesssim 2$, where the ESTM predictions have been validated with GCM experiments (see Vladilo et al. 2015). In this range, the ESTM predicts modest changes of the global temperature of Kepler-452b (middle row of Fig. \ref{figEccRotObl}, left panel). On the other hand, the equator-pole gradient decreases significantly with increasing $P_\mathrm{rot}$ (central panel) due to the reduction of Coriolis forces and the consequent rise of meridional transport efficiency. The resulting impact on the habitability index $h_{050}$ is shown in the right panel. For the models with low CO$_2$ content (green and blue curves) the rise of the equator-to-pole transport efficiency cools the tropical regions and even increases the habitability. For the model with high CO$_2$ content (red curves), the same mechanism 
warms the polar regions but is unable to effectively cool the tropics, and a large fraction of the planet surface warms above  $T \sim 50^\circ$C. In this case, the rise of meridional transport decreases the already low habitability.  

Variations of obliquity of the planet rotation axis, $\epsilon$, are well known to influence the climate and have not been investigated so far for Kepler-452b. The axis tilt determines the seasonal and latitudinal insolation of the surface and, as a result, it governs the strength and direction of the temperature gradients which drive the atmospheric circulation. At low axis tilts, the temperature gradient drives the circulation from the tropics to the poles, as in the case of the Earth. When the tilt is high, the seasonal changes generate an evolving pattern of temperature gradients that can only be tracked with 3D climate models (e.g. Ferreira et al. 2014). The ESTM calculates exactly the seasonal and latitudinal insolation, but is able to simulate the atmospheric transport only for moderate values of the axis tilt. The bottom panels of Fig. \ref{figEccRotObl} show the impact of variations of the axis obliquity in the range $0^\circ \leq \epsilon \leq 45^\circ$ where the ESTM has been validated with 3D climate experiments. In the left panel one can see that the impact on the mean global temperature is relatively small. However, the equator-pole temperature gradients decreases significantly with increasing $\epsilon$ (central panel) because at high axis tilt the insolation has a very different latitudinal distribution at different phases of the orbital period. The resulting habitability is shown in the right panel. For the models with low CO$_2$ content (green and blue curves) the more even latitudinal distribution of insolation prevents eccessive heating in any specific zone and increases the habitability. For the model with high CO$_2$ content (red curves), the broader distribution of insolation warms above $T \sim 50^\circ$C the polar zones which would be  cooler at  $\epsilon\sim 0^\circ $. In this case, an increase of axis tilt decreases the already low habitability.   

In addition to the above planetary parameters, we also explored the impact of variations of the ocean/continental coverage on the habitability of Kepler-452b. In the coupled atmosphere-ocean GCM simulations of Hu et al. (2017) a ocean coverage of 100\% was adopted. In the ESTM we can define a schematic geography by changing the fraction of oceans and continents in each latitude zone. At variance with the coupled atmosphere-ocean GCM, the ESTM does not include the ocean transport, but it does take into account differences in albedo, thermal capacity and cloud cover between oceans and continents. For simplicity, we adopted a constant fraction of oceans, $f_o$, and continents, $1-f_o$, in each latitude zone and we explored the effect of varying the ocean coverage in the range $0.1 \leq f_o \leq 1$. We found a modest increase of $<\!T\!>$  and very small changes of $\Delta T_\mathrm{ep}$ with increasing $f_o$. In the models with minimal CO$_2$ content, the ice cover shows a moderate decrease with increasing $f_o$, leading to a weak rise of habitability. In the models with Earth-like and higher CO$_2$ content, the ice cover is negligible and the habitability shows a moderate decrease with increasing $f_o$. To understand how the ocean transport would modify these results, a series of atmospheric-ocean GCM simulations with different geographies should be performed. 
 

\begin{figure*}     
\begin{center}
\includegraphics[width=\linewidth]{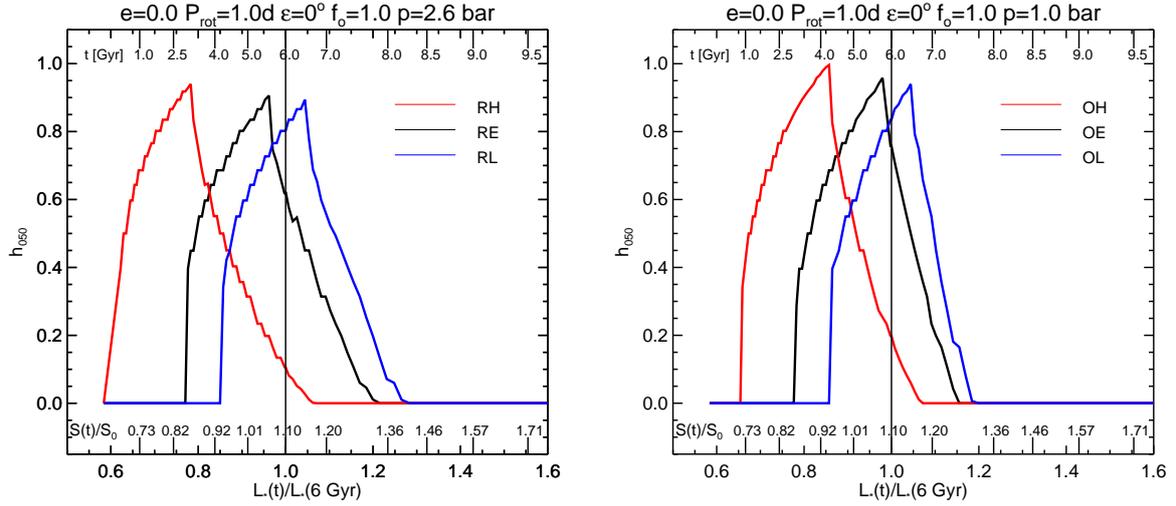} 
\caption{Effect of the CO$_2$ abundace on the evolution of the habitability index $h_{050}$.  $h_{050}$ is shown as a function of the stellar luminosity $L_\star(t)/L_{\star}(6\,Gyr)$, of the corresponding insolation relative to solar $S(t)/S_0$ (reported for e=0), and of the stellar age in Gyr. Left panel: red, black and blue lines are for RH, RE, RL respectively as defined in Table \ref{tabModels}. Right panel: red, black and blue lines are for OH, OE, OL respectively  as defined in Table \ref{tabModels}. See text for details.
}
\label{figHabEvol2}
\end{center}
\end{figure*} 

\begin{figure*}     
\begin{center}
\includegraphics[width=\linewidth]{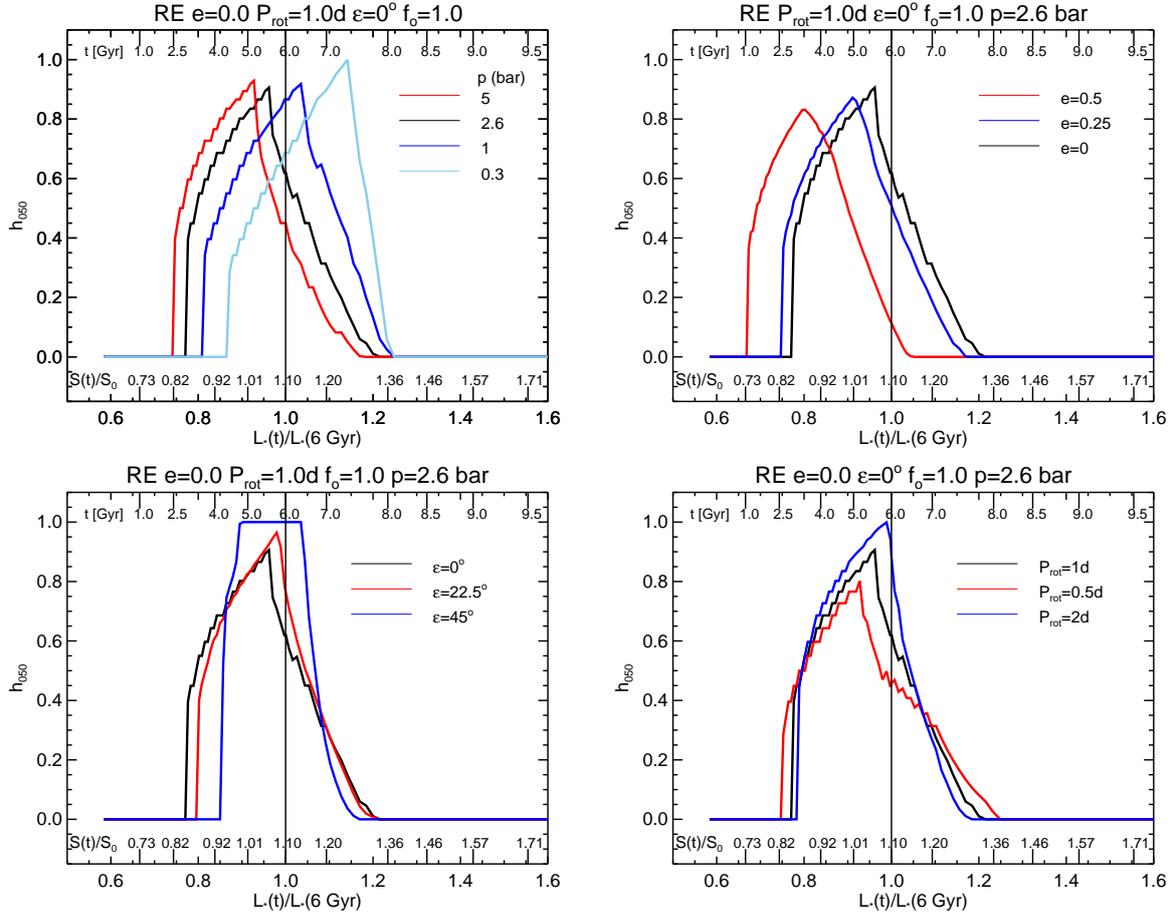} 
\caption{Same as in Fig.\ \ref{figHabEvol2} but for the RE model only, by varying one parameter per time. In each plot the black line refer to the RE model as defined in Table \ref{tabModels}.   
Top left panel: effect of pressure, p=5 bar (red), p=2.6 bar (black), p=1 bar (blue), p=0.3 bar (light blue). 
Top right panel: effect of eccentricity, e=0 (black), e=0.25 (blue), e=0.5 (red).
The insolation values $S/S_0$ reported for e=0 scale as $(1-e^2)^{-0.5}$. At 6 Gyr $S/S_0$=1.27 and 1.14 respectively for e=0.5 and 0.25. 
Bottom left panel: effect of obliquity, $\epsilon=0^\circ$ (black), $22.5^\circ$ (red) and $45^\circ$ (blue). 
Bottom right panel: effect of rotation period, $P_{rot}=1$ d (black), $0.5$ d (red) and 2 d (blue).
}
\label{figHabEvol4}
\end{center}
\end{figure*} 

\begin{figure*}     
\begin{center}
\includegraphics[width=\linewidth]{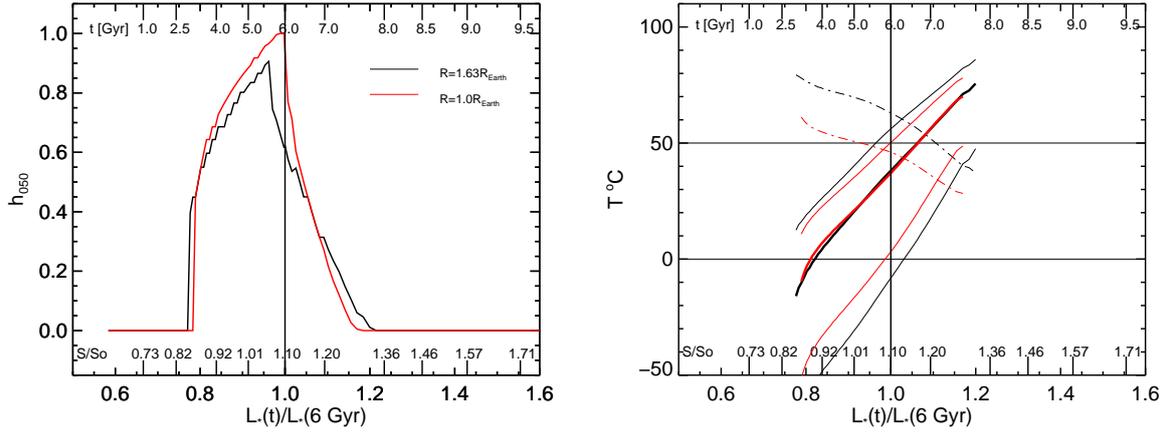} 
\caption{Effect of planetary radius on the temperature and habitability for the RE reference model of Table \ref{tabModels}. Black line: R$=1.63$R$_\oplus$. Red line: R=$1.0$R$_\oplus$. Left panel: $h_{050}$ habitability. Right panel: mean global T (thick solid lines); minimum and maximum T reached in any latitude strip  along the orbid (thin solid lines); mean equator to polar T difference (dot-dashed lines).}
\label{figcfrR}
\end{center}
\end{figure*}

\begin{figure*}     
\begin{center}
\includegraphics[width=\linewidth]{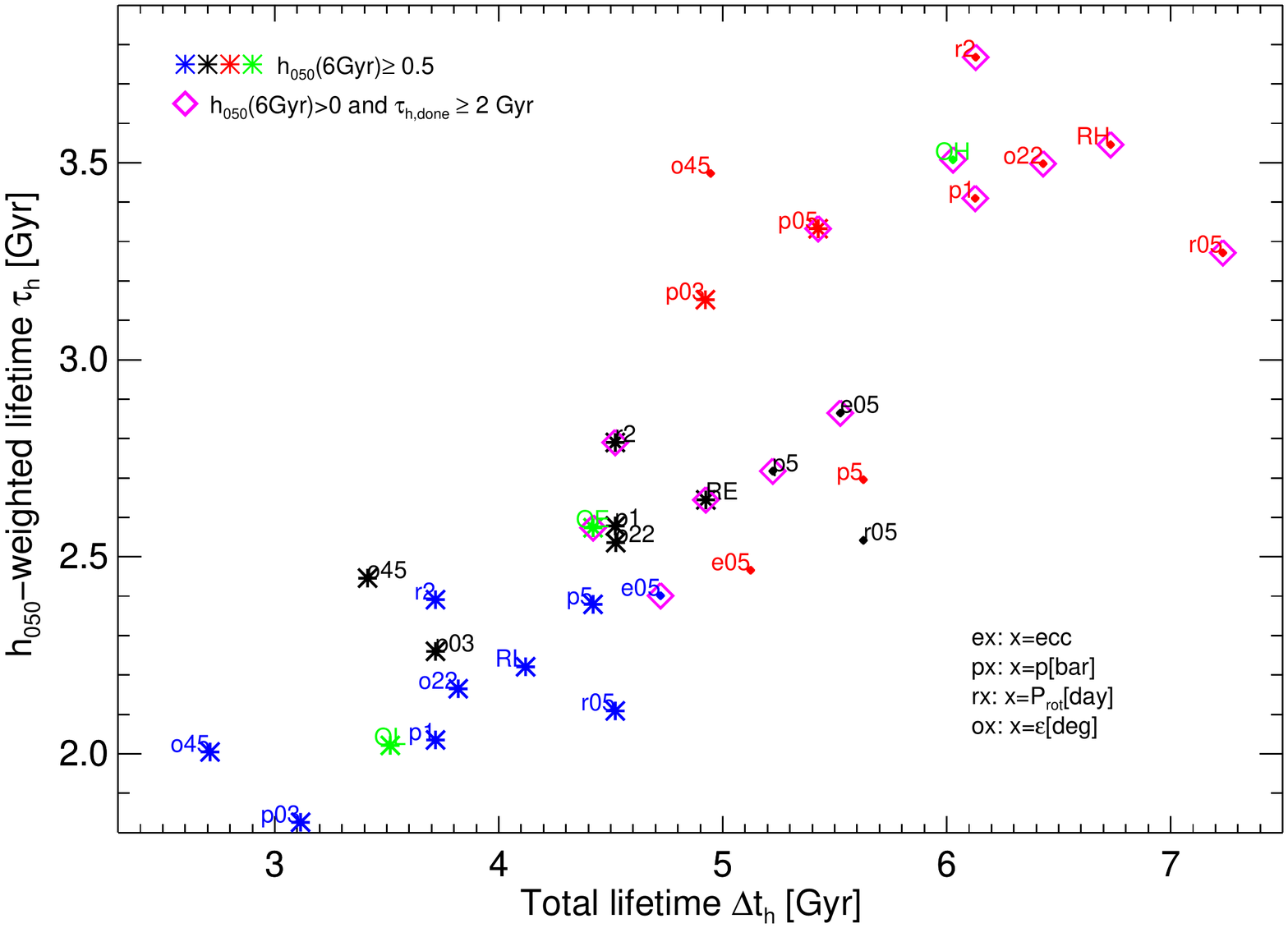} 
\caption{Total habitability-weighted time versus total span of time for which $h_{050} > 0$. The reference models as defined in Table \ref{tabModels} are indicated with their names. The red, black, and blue colors refer respectively to the high-, earth-like, and low-CO$_2$ content for the rock case (see Table \ref{tabModels}). Instead the Ocean-rock planet case is in green for the 3 compositions.
The labels of the other cases refer to that single parameter varied with respect to the reference models of Table \ref{tabModels}. The parameter is identified by its initial name (e=eccentricity, p=pressure, r=rotation period, o=obliquity) and its value. 
For clarity the intermediate e=0.25 cases are not shown, and the p=0.5 bar case is shown only for the red models. 
The asterisks highlight those models which currently (6 Gyr) have $h_{050} \geq 0.5$. 
Those models that currently have still $h_{050}>0$ and have had at least  $\tau_{h,done} \geq 2$ Gyr since the start of their HZ are highlighed by a magenta romb. These are the most interesting cases for the potential presence of atmospheric biosignatures. 
}
\label{figHabTime}
\end{center}
\end{figure*}

\subsection{Evolution of the planet habitability}\label{secEvoHab}

The surface habitability of a planet evolves on the long term as a function of planetary, astronomical and possibly biological factors. The potential generation of detectable atmospheric biosignatures requires long-term climate stability and habitability. A basic estimate of the habitability timescale can be calculated from the time spent by the planet in the HZ
during the main-sequence luminosity evolution of the host star, assuming non-evolving orbital factors.
The width of the classical HZ is computed by assuming an evolving atmospheric composition, with the inner edge defined by the onset of a runaway or moist greenhouse instability in a water-vapor dominated atmosphere, and the outer edge defined by the maximum greenhouse effect allowed by a CO$_2$ dominated atmosphere (e.g. Kasting et al. 1993; William \& Kasting 1997; Kopparapu et al. 2013). The latter limit rests on the assumption that a carbon-silicate geological negative climate feedback acts on exoplanets as it does on Earth to regulate the amount of CO$_2$ as a function of surface temperature (Walker, Hays \& Kasting 1981; Kasting et al.\ 1993). 
However, the development and long-term maintainance of a geologic activity on exoplanets of different masses (e.g. O\textquotesingle Neill \& Lenardic 2007; Valencia, O\textquotesingle Connell \& Sasselov 2007; Korenaga 2010; Foley, Bercovici \& William 2012; Stein, Lowman \& Hansen 2013; Noack \& Breuer 2014), or the efficacy of the cycle as a climate thermostat (e.g. Kump, Brantley, \& Arthur 2000; Menou 2015; Turbet et al. 2017) are debated. 

In our work we use a stellar luminosity track to account for the luminosity evolution of the host star and we estimate
the impact of variations of the atmospheric composition using the three values of CO$_2$ defined in Table \ref{tabModels}. 
For each value of CO$_2$ we calculate the evolution of the habitability index $h_{050}$ as a function of planet insolation.
The habitability time span is calculated from the range of insolations which yield $h_{050}>0$. As explained in Section \ref{secMethod}, the relatively low maximum temperature allowed by this biosignature-optimized index allow to avoid the uncertainties inherent to the onset of water-vapor driven climate instabilities, which our model can not track. 
The minimum and maximum insolation values within which $h_{050}>0$ provide a maximum habitability time span, $\Delta t_h=t^\star (S_{max})-t^\star (S_{min})$. Since within this time interval the habitable surface fraction of the planet is not constant, we also calculate an effective habitability time obtained by weighting each time step by the corresponding value of $h_{050}$: $\tau_h = \sum h_j \, \delta t^\star_j$, where $\delta t^\star_j$ is the stellar age increment for a given insolation increment (see Silva et al.\ 2016). 
In practice, by providing larger weight to stellar evolutionary phases corresponding to larger surface habitality fractions, $\tau_h$  provides a timespan more 
representative from the habitability point of view. 
We may expect that in a habitable $h_{050} >0$ condition, but involving a small surface fraction even for a total time of a few Gyr, a widespread production of biosignatures would be difficult. By setting a minimum value for $\tau_h$ as a reference for the potential production of biosignatures (2 Gyr, see below), we are also imposing a minimum average habitable surface fraction during the total lifetime $\Delta t_h$ ($\gtrsim$ 30\% with the range of values $\Delta t_h \sim $ 3 to 7 Gyr found for our models, see below and Fig.\ \ref{figHabTime}).

To select the stellar luminosity track, we adopted the properties of the host star Kepler-452 (KIC 8311864), a G2 main sequence star, as determined by Jenkins et al.\ (2015) by spectroscopic observations and stellar models. The faintness of the star did not allow direct mass, radius and age measurements via asteroseismology.   
By taking as a reference the current luminosity of the star, $20\%$ higher than the solar luminosity, its estimated current age of 6 Gyr and a metallicity $\sim 60\%$ higher than solar (values in table 2 of Jenkins et al.\ 2015), we have adopted from the PARSEC\footnote{http://people.sissa.it/$\sim$sbressan/\texttt{CAF09\_V1.2S\_M36\_LT}} 
models (Bressan et al.\ 2012) a stellar track with M$_\star$=1.05 M$_\odot$ and $Z=0.03$.
These stellar models include also the pre-MS evolution, a phase characterized by fast variations of the stellar luminosity and possibly of the orbital and planetary parameters, including volatiles accretion. We consider the luminosity evolution of the star since its first arrival on the MS, when also planetary dynamical stability may be established, and variations of the atmospheric mass may be considered mild if a solid crust has formed and atmospheric loss can be neglected.
Since each evolutionary curve of habitability is calculated at constant CO$_2$
our estimates of  $\Delta t_h$ and $\tau_h$ should provide lower limits to the habitability timescales
compared to the case in which a CO$_2$ stabilizing feedback is at work. 

The evolutionary track that we have adoped enters the MS at an age of  $t_i=0.045$ Gyr with an initial minimum luminosity $L_{\star}(t_i)/L_{\star}(6 Gyr)=0.58$. The habitability evolution is computed as a consequence of the increase of the insolation received by the planet as the star ages along the MS:
$S(t)/S(6 Gyr)=L_{\star}(t)/L_{\star}(6 Gyr)$, with $L_{\star}(t)/L_{\star}(6 Gyr)$ given by the track, and the current insolation with respect to solar given by $S(6 Gyr)/S_\odot=(L_{\star}(6 Gyr)/L_\odot)/a^2/\sqrt{(1-e^2)}$, 10\% higher than solar for zero eccentricity.

In Fig.\ \ref{figHabEvol2} we show the evolution of $h_{050}$ for our reference rocky and ocean models as defined in Table \ref{tabModels}, for the 3 CO$_2$ abundances, as a function of the stellar luminosity $L_\star(t)/L_{\star}(6\,Gyr)$, of the planetary insolation relative to solar $S(t)/S_\odot$ for e=0, and of the age of the star. 
In both panels of Fig.\ \ref{figHabEvol2} a similar trend of the habitability as a function of the insolation and of the CO$_2$ abundance is obtained. As expected, the insolation and age values for which $h_{050} > 0$ are a strong function of the atmospheric composition. The larger greenhouse effect of the RH and OH models implies that the planet would currently be almost out of the HZ (red lines), unless in the meantime the CO$_2$ abundance had decreased. In fact, for the rocky case, the total $\Delta t_h$ would be $\gtrsim 8$ Gyr if the CO$_2$ amount were to adjust at increasing insolation within the range considered in Fig.\ \ref{figHabEvol2}. This extended total lifetime is close to the one estimated by Jenkins et al. (2015).   
According to the E (black lines) and L (blue lines) models the planet would currently be still habitable, with $\sim 62$\% and $76$\% habitable surface respectively for RE and OE in their declining phases, while for RL and OL the peak value of the habitability would have not yet been reached. In all cases, even though the widths of the HZ are comparable in insolaton ($\Delta S \sim 0.4-0.5$) the correspondig extention in time is maximized for early habitabilities (discussed below).

In Fig.\ \ref{figHabEvol4} the effect  of the different parameters on the habitability of the model RE is shown. In each panel one single parameter per time has been varied. As shown in Figs. \ref{figPressure} and \ref{figEccRotObl}, the variations of pressure and eccentricity heavily impacs on both the surface temperature and its gradient. Because of this, they determine a net shift of the HZ in insolation/time. On the other hand, different obliquities and rotation periods, within the range of values we can consider in our model, imply a larger effect on the surface energy distribution than on the absolute mean temperature. As a result, they cause an important modulation of the habitability curve but not net shifts. Particularly important is the flattening of the surface temperature gradient obtained at $45^\circ$ obliquity (Fig. \ref{figEccRotObl} bottom central panel) that allows an extended phase of constant 100\% habitability (blue curve, bottom left panel of Fig.\ \ref{figHabEvol4}). A more efficient  surface energy redistribution is obtained also by increasing the planet rotation period (Fig.\ \ref{figEccRotObl} middle panel). The increase of $P_{rot}$ from 0.5 to 2 days (bottom right panel of Fig.\ \ref{figHabEvol4}), that is the maximum value we can reliably consider with our model, produces a general increase of the habitability curve. 

In all other cases, the habitability results peaked at an optimum insolation at which the surface temperature gradient has reached a minimum value (due to the increasing moist transport for increasing insolation) before the equatorial latitude zone overcomes the maximum allowed temperature of $50^\circ$C, at this point the equatorial habitability drops to 0, causing the sharp drop of $h_{050}$ after the maximum.
The habitability depends on the steepness of the latitude temperature profile and temperature extreme values with respect to the temperature range chosen as the habitability criterium. In Figs. \ref{figPressure} and \ref{figEccRotObl} it can be seen that most models have $\Delta T_\mathrm{ep} > 50^\circ$C (and generally less steep but for very hot models), for this reason a $100\%$ habitability is almost never reached, and the habitability curve is peaked unless a very efficient surface energy redistribution is at work (e.g. at high inclinations). 
As discussed in Section \ref{secCompEarth}, a large planetary radius increases the equatorial to polar temperature differences that are partly counterbalanced by the opposite effect of an increasing insolation. But at the same time the latter causes maximum temperatures higher than the allowed limit before smoothing the temperature profile enough for an extended period of high habitability.  
As an explicit example we show in Fig. \ref{figcfrR} the effect on temperature and habitability that a smaller radius would have on the reference RE model. The left panel shows the $h_{050}$ index vs insolation for the RE model (in black) and the same model but with R=$1.0$R$_\oplus$ (in red). In the right panel the corresponding mean global temperatures (thick lines), minimum and maximum planet temperatures reached in any latitude strip alog the orbit (thin lines), and $\Delta T_\mathrm{ep}$ (dot-dashed lines) are shown. The different efficiency of the meridional energy distribution has a minor effect on $<T>$, but heavily affects the latitudinal temperature profile and the temperature extremes, and the interplay of these quantities determines the value and extent of the habitability.

A summary of the time extent of the habitability as a function of the different models discussed so far is shown in Fig.\ \ref{figHabTime}. We plot the total $h_{050}$-weighted time $\tau_h$ versus the total time span $\Delta\,t_h$ for the reference models, labeled with their names on the plot. For the rocky planet case, we label each model according to that single parameter that has been varied from the RL, RE or RH case reported in Table \ref{tabModels}, with the color indicating the CO$_2$ abundance (respectively blue, black and red for L, E and H). The three O models (all in green) are shown only for the reference choice of parameters. 
Since the rate of the stellar luminosity evolution increases with age, for a given width of the HZ in insolation, the corresponding habitability time span depends on when, during the stellar evolution, the right conditions are met (e.g., for the track we have adopted, $dL/dt$ $\sim 0.05$, $0.07$, $0.11$, $0.2$ $L_\odot/$Gyr at  $t=1$, 3, 6, 8 Gyr respectively).
As a result, those solutions that imply earlier habitability in the stellar evolution are characterized by longer total lifetime spans.
This is evident from the general increase of both $\Delta\,t_h$ and $\tau_h$  from the L to the E and H composition. This same trend is found for increasing $p$ and $e$  at a given composition. The drop for the RH cases with $p=5$ bar and $e=0.5$ is because the HZ would extend to luminosities lower than the minimum luminosity of the star at the start of the MS. 
Within a given composition, the trend is reversed for increasing rotation period, so that a slower rotating planet implies higher effective $\tau_h$ due to the larger habitability in spite of the shorter total time span. This is more evident with the different values of the inclination.   
Quite large differences in $\Delta\,t_h$ are in all cases obtained by increasing the obliquity, but with a not so marked difference on the corresponding weighted time for the 3 obliquity values due to the much larger values of the habitable surface fraction reached in particular at $45^\circ$.

The asteriscs highlight those models for which currently the planet would have more than 50\% of its surface habitable. This excludes all the high-CO$_2$ cases, except at low pressure ($p \leq 0.5$ bar). The magenta rombs highlight those models that currently have still $h_{050}>0$ and for which the effective habitability time has been, since the start of the HZ up to the current age of 6 Gyr, $\tau_{h,done} \ge 2$ Gyr. In Silva et al. (2016) we took 2 Gyr as a reference lower limit for the potential development of a widespread biosignature-producing biological activity. 
In principle, even for those cases currently on the verge of exiting their habitable phase, potentially still detectable atmospheric biosignatures could be possible. 
All models with low CO$_2$ content, except the RL cases with $p \leq 0.5$ bar, have a total $2 < \tau_h < 2.5$ Gyr, but there would not have been enough time yet to pollute the atmosphere. Instead, due to the late start of the right conditions, for these models a relatively long residual lifetime ($\tau_{h,res} \geq 1$ Gyr) would be possible. Our reference cases RE and OE with a residual effective time $\tau_{h,res}$ shorter than 1 Gyr, would anyway be in a promising condition for the potential presence of atmospheric biosignatures.

\section{Discussion and conclusions}\label{secConc}

Currently Kepler-452b is the best available example of a terrestrial-size and probably rocky planet within the habitable zone of a solar-type star. 
Several works exploring the frequency of planets in the Galaxy, generally agree on an expected common occurence of Earth-like rocky planets around solar and later type stars (e.g., Traub 2012; Dressing \& Charbonneau 2013; Fressin et al.\ 2013; Petigura, Howard \& Marcy 2013; 
Batalha 2014; Foreman-Mackey, Hogg \& Morton 2014; Silburt, Gaidos \& Wu 2015). This expectation is further supported by theoretical planet formation models (e.g. Mordasini et al. 2015 for a thorough review). 
It is therefore predictable that future observational facilities will increase the detection of rocky planets in the HZ of not only M-type but also of solar-type stars, even if a deep characterization of long period rocky planets will probably have to wait for dedicated missions such as PLATO, whose targets should be accessible by the next generation large telescopes (Rauer et al. 2014).

The present study is a test case for planets whose number is expected to increase in the near future, but whose characterization needs to await longer times. For these planets, it is important to highlight the most promising targets in terms of potential presence of atmospheric biosignatures.
To this aim, we have explored the surface fractional habitability of Kepler-452b with a temperature-dependent index designed to maximize the potential presence of biosignature-producing biological activity (Silva et al.\ 2016). We have computed the surface temperature with our climate model (ESTM, Vladilo et al. 2015), as a function of the known planeray parameters as measured by Jenkins et al.\ (2015), and of the yet unknown parameters. 
The measured quantities for Kepler-452b do not currently include its mass. The probability of a pure rocky composition for this 
planet has been quantified by Jenkins et al. to lie between 49 and 62\%. At lower mean densities, ice-rock mixtures could extend the probability of a rocky- rather than a gaseous-dominated planet. But at this stage the rocky nature, at the base of the habitability studies, has to be assumed.

We have first validated our model by comparing with the GCM computations for Kepler-452b by Hu et al.\ (2017). We have then explored the habitability of Kepler-452b for 2 values of surface gravity (g=1.6 and 1.0 g$_\oplus$), 3 values of CO$_2$ abundances (earth-like, 100 times higher and  $\sim 3\%$ the earth value), and a range of values of pressure, eccentricity, rotation period, axis obliquity and ocean surface fraction. Most of these parameters have been explored here for the first time.

The main results of our parameter exploration of the present day habitability are: 
(i) most combinations of the parameters yield habitable solutions for Kepler-452b, particularly for an Earth- or lower CO$_2$ abundance. For the high CO$_2$ abundance case, the surface pressure should be $\lesssim 1-2$ bar to allow $\gtrsim 20\%$  habitability, and $\lesssim 0.5$ bar for $\geq 50\%$ habitability. For lower CO$_2$ content, most solutions have currently $\geq 50\%$ habitable surface; 
(ii) the strongest effect on surface temperature are due to atmospheric composition, surface pressure, and eccentricity. However even those parameters that, within the range we have explored, do not imply significant variations of the mean global temperature (e.g., $P_{rot}$, $\epsilon$ and $f_o$) give rise to extremely different values of the habitable surface; 
(iii) therefore, the mean global temperature may be taken just as an indication of the general planetary surface condition, the habitability being dependent on the details of the temperature latitudinal gradient and its orbital evolution. This means that the habitability depends on if and when the extreme values of temperatures are within or out of the temperature range of the adopted habitability criterium. All the considered parameters concur to determine the exact value of the habitability;
(iv) the large planetary radius, that damps the meridional energy transport, tends to steepen the latitudinal temperature gradient above our adopted habitable temperature range. This is partly offset by the opposite effect caused by a relatively high insolation (that rises the efficiency of the moist transport), determining for most cases a quite large current habitability. 

Since atmospheric biosignatures may conceivably be produced in detectable amounts only in presence of a widespread and long term surface biological activity, we have also estimated the effect of the stellar luminosity evolution on the habitability lifetime span.
This provides a basic and lower limit lifetime, if geological negative feedbacks stabilize the climate. 
The curves of habitability evolution that we find are in most cases very steep and with a maximum habitability $< 1$ due to the equatorial temperature overcoming the maximum allowed temperature before the latitude temperature profile can be sufficiently smoothed by the increasing insolation.  
The habitability time span is maximized early in the stellar evolution due to the increase in time of the rate of the stellar luminosity evolution. For all models (except the low CO$_2$ and low pressure case), the total lifetime is $\geq 2$ Gyr (in terms of the habitability-weighted time we have adopted as a more meaningful quantity), that we take as a representative minimum value for the development of widespread biological activity.
All high-CO$_2$ models that currently are on the verge of exiting their habitable phase would have had enough time to potentially produce atmospheric bisognatures. This holds also for our reference models with earth-like composition (RE and OE models) that in addition would currently have high habitability ($\geq 50\%$), but with a residual lifetime $\lesssim 1$ Gyr. High current habitabilities with long residual lifetimes are found for most low-CO2 models, but in this case with not enough elapsed time for a potential atmospheric pollution. For our reference rocky model, if the CO$_2$ amount is allowed to evolve within our explored range, the total non-weighted lifetime would be $\gtrsim 8$ Gyr.  
Similarly, the presence of an active biosphere could counterbalance the effects of the changes in the stellar luminosity and prolong the habitability period if a negative feedback mechanism is in place, such as that conceptualized in the parable of Daisyworld by Watson \& Lovelock (1983). Such possibility will be explored in a future work. 

All these results depend on the choice of a value of mass and surface gravity. 
It can be expected that these quantities and the eccentricity will become measurable in the future for habitable planets around solar type stars. The largest uncertainties will then be the atmospheric properties. We have shown that the latter, together with the other unknown planetary parameters have to be taken into account to quantify the potential surface habitability of exoplanets.

 
\section*{Acknowledgements}

We are grateful to Prof. Yongyun Hu for providing us details of his GCM model for Kepler-452b.


\bsp	
\label{lastpage}

\begin{thebibliography}{99}

\bibitem[]{}Batalha N. M., 2014, PNAS, 111, 12647

\bibitem[]{}Bressan A., Marigo P., Girardi L., Salasnich B., Dal Cero C., Rubele S., Nanni A., 2012, \mnras, 427, 127

\bibitem[]{}Caldeira K., Kasting J.F., 1992, \nat, 360, 721

\bibitem[]{}Clarke A., 2014, IJAsB, 13, 141

\bibitem[]{}Dressing C.~D., Charbonneau D., 2013, \apj, 767, 95

\bibitem[]{}Ferreira D., Marshall J., O\' Gorman P. A., Seager S., 2014, \icarus, 243, 236

\bibitem[]{}Foley B. J., Bercovici D.,  Landuyt W., 2012, E\&PSL, 331, 281

\bibitem[]{}Foreman-Mackey D., Hogg D. W., Morton T. D., 2014, \apj, 795, 64
	
\bibitem[]{}Fortney J. J., Marley M. S., Barnes J. W., 2007, \apj, 659, 1661

\bibitem[]{}Fressin F., Torres G., Charbonneau D., Bryson S. T., Christiansen J., Dressing C. D., Jenkins J. M., 
Walkowicz L. M., Batalha N. M.,  2013, \apj, 766, 81

\bibitem[]{}G{\"u}del M., Dvorak R., Erkaev N., Kasting J., Khodachenko M., Lammer H., Pilat-Lohinger E., Rauer H., Ribas I., Wood B.~E., 2014, Protostars and Planets VI, H. Beuther, R. S. Klessen, C. P. Dullemond, and T. Henning (eds.), University of Arizona Press, Tucson, p.883-906

\bibitem[]{}Jenkins J. M. et al., 2015, \aj, 150, 56 

\bibitem[]{}Hu Y., Wang Y., Liu Y., Yang J., 2017, \apjl, 835, 6

\bibitem[]{}Kaspi Y., Showman A. P., 2015, \apj, 804, 60

\bibitem[]{}Kasting J. F., 1988, \icarus, 74, 472

\bibitem[]{}Kasting, J. F., Whitmire D. P., Reynolds, R. T., 1993, \icarus, 101, 108

\bibitem[]{}Kasting J. F., Kopparapu R., Ramirez R. M., Harman C. E., 2014, PNAS, 111, 12641

\bibitem[]{}Kiehl J. T., Hack J. J., Bonan G. B., et al., 1998, JCli, 11, 1131

\bibitem[]{}Kopparapu R. K., Ramirez R., Kasting J. F., Eymet V., Robinson T. D., Mahadevan S., Terrien R. C., Domagal-Goldman S., Meadows V., Deshpande R., 2013, \apj, 765, 131

\bibitem[]{}Kopparapu, R. K., Ramirez R. M., SchottelKotte J., Kasting J. F., Domagal-Goldman S., Eymet, V., 2014, \apjl, 787, 29

\bibitem[]{}Korenaga J., 2010, \apjl, 725, 43

\bibitem[]{}Kump L.R., Brantley S. L., Arthur M. A., 2000, Annu. Rev. Earth Planet. Sci., 28, 611

\bibitem[]{}Leconte J., Forget F., Charnay B., Wordsworth R., Pottier A., 2013, \nat, 504, 268

\bibitem[]{}Lopez E. D.,  Fortney J. J., 2014, \apj, 792, 1

\bibitem[]{}Marcy G. W., Weiss L. M., Petigura E. A., Isaacson H., Howard A. W., Buchhave L. A., 2014, PNAS, 111, 12655

\bibitem[]{}Menou K., 2015, Earth and Planetary Science Letters, 429, 20

\bibitem[]{}Mordasini C., Molli\`ere P., Dittkrist K.-M., Jin S., Alibert Y., 2015, IJAsB, 14, 201

\bibitem[]{}Noack L., Breuer D., 2014, \planss, 98, 41

\bibitem[]{}North G. R., Coakley J. A., 1979, JAtS, 36, 1189
\bibitem[]{}North G. R., Mengel J. G., Short D. A., 1983, JGR, 88, 6576

\bibitem[]{}O\textquotesingle Neill C., Lenardic A., 2007, \grl, 34, 19204

\bibitem[]{}Petigura E. A., Howard A. W., Marcy G. W., 2013, PNAS, 110, 19273

\bibitem[]{}Precht H., Christophersen J., Hensel H., Larcher W., 1973, Temperature
and Life. Springer-Verlag, Berlin, Heidelberg

\bibitem[]{}Rauer H. et al., 2014, Experimental Astronomy, 38, 249


\bibitem[]{}Rogers L. A., 2015, \apj, 801, 41

\bibitem[]{}Rowe J. F., et al., 2014, \apj, 784, 45

\bibitem[]{}Seager S., 2013, \sci, 340, 577
\bibitem[]{}Seager S., 2014, PNAS, 111, 12634

\bibitem[]{}Shields A. L., Ballard S., Johnson J. A., 2016, \physrep, 663, 1

\bibitem[]{}Silburt A., Gaidos E., Wu Y., 2015, \apj, 799, 180

\bibitem[]{}Silva L., Vladilo G., Schulte P., Murante G., Provenzale A., 2016, IJAsB, doi:10.1017/S1473550416000215, arXiv:1604.08864  

\bibitem[]{}Snellen I. A. G., Brandl B. R., de Kok R. J., Brogi M., Birkby J., Schwarz H., 2014, \nat, 509, 63

\bibitem[]{}Spiegel D. S., Menou K., Sharf C. A., 2008, \apj, 681, 1609

\bibitem[]{}Spiegel D. S., Menou K., Sharf C. A., 2009, \apj, 691, 596

\bibitem[]{}Stein C., Lowman J. P., Hansen U., 2013, E\&PSL, 361, 448

\bibitem[]{} Tinetti G., Encrenaz T., Coustenis A., 2013, \aapr, 21, 63

\bibitem[]{}Traub W. A., 2012, \apj, 745, 20 

\bibitem[]{}Turbet M., Forget F., Leconte J., Charnay B., Tobie, G., 2017, arXiv:1703.04624

\bibitem[]{}Valencia D., O\textquotesingle Connell R. J., Sasselov D. D., 2007, \apjl, 670, 45

\bibitem[]{}Vladilo G., Murante G., Silva L., Provenzale A., Ferri G., Ragazzini G., 2013, \apj, 767, 65
\bibitem[]{}Vladilo G., Silva L., Murante G., Filippi L., Provenzale A., 2015, \apj, 804, 50

\bibitem[]{}Walker J. C. G., Hays P. B., Kasting J. F., 1981, \jgr, 86, 9776

\bibitem[]{}Watson A.J., Lovelock  J. E., 1983, Tellus, 35B, 284-289

\bibitem[]{}Williams D. M., Kasting J. F., 1997, \icarus, 129, 254

\bibitem[]{}Xie J.-W., Dong S., Zhu Z., Huber D., Zheng Z., De Cat P., Fu J., Liu H.-G., Luo A., Wu Y., Zhang H., Zhang H., Zhou J.-L., Cao Z., Hou Y., Wang Y., Zhang Y., 2016, PNAS, 113, 11431

\end{thebibliography}
\end{document}